\documentclass[preprint,showpacs,preprintnumbers,amsmath,amssymb,showkeys]{revtex4}

\usepackage{graphicx}
\usepackage{bm}

\begin{document}

\preprint{}

\title{Collective modes of a quasi two-dimensional Bose condensate in large gas parameter regime}

\author{S R Mishra}
\email{srm@cat.ernet.in}
\author{S P Ram}
\author{Arup Banerjee}
\affiliation{Laser Physics Application Section, Raja Ramanna Centre for Advanced Technology, \\Indore 452013, India.}

\begin{abstract}
 We have theoretically studied the collective modes of a quasi two-dimensional (Q2D) Bose condensate in the large gas parameter regime by using a formalism which treats the interaction energy beyond the mean-field approximation. In the calculation we use the perturbative expansion for the interaction energy by incorporating the Lee, Huang and Yang ( LHY) correction term. The results show that incorporation of this higher order term leads to detectable modifications in the mode frequencies.
\end{abstract}

\pacs{\\
03.75.Kk Dynamic properties of condensates; collective and hydrodynamic excitations, superfluid flow.\\
03.75.Hh Static properties of condensates; thermodynamical, statistical, and structural properties.\\    
05.30.Jp Boson systems (for static and dynamic properties of Bose-Einstein condensates). }
\keywords{Bose-Einstein condensation, collective modes, scattering length.}
\maketitle

\section{Introduction}
\label{intro}
Bose-Einstein Condensation (BEC) of dilute atomic gases has been achieved in a variety of atomic samples using different magnetic and optical traps ~\cite{cornell,ketterle,hulet_1995,hulet_1997,cennini}. These advances have resulted in a investigations of properties of these ultra-cold gases in different conditions. At present, there is considerable interest in BEC in low dimensions where particle motion is restricted in one or more dimensions ~\cite{arup1,smith,heche,kam,ghosh,ho,tanatar,gorlitz,pitaev}. This is because a Bose gas confined in low dimensions shows remarkably different behaviour than it shows in three dimensions. For example, a two-dimensional homogeneous Bose gas does not undergo Bose-Einstein condensation on cooling. Instead, there is a phase transition called Berezinski-Kosterlitz-Thouless  transition in which system becomes superfluid without having long-range ordering. For a homogeneous Bose gas confined in one dimension, BEC does not exist even at absolute zero. However, in inhomogeneous confining potentials BEC can exist in lower dimensions also ~\cite{bagnato}.
 
 Recently, several groups have experimentally investigated the properties of Bose condensates confined in the low dimensions ~\cite{smith,heche,kam}. To investigate the macroscopic quantum behaviour of these systems, the study of collective modes has proved to be a fundamental tool. Usually, the theoretical description of these dynamical properties such as collective modes of a Bose condensate can be accomplished by Gross-Pitaevskii (GP) equation taking interaction energy within the mean-field approximation. However, it is known that the mean-field theory gives accurate results only in the regime of small values of gas parameter($ \rho a_{s}^{3},\rho $ is density and $a_{s}$ is scattering length). With the recent experimental advances in this field, access to the large gas parameter regime has become feasible either by increasing the number of atoms in the condensate or by increasing the scattering length via Feshbach resonances. Therefore it is essential to assess the GP theory in this regime. In the present theoretical studies, we have calculated the collective mode frequencies of a Q2D gas in the regime of large gas parameter by using a perturbative expansion for the interaction energy.  We have incorporated the Lee, Huang and Yang ( LHY) correction term in the energy expansion to go beyond the mean-field approximation ~\cite{arup2}. We find that detectable changes in the collective mode frequencies can occur when we incorpoarte the LHY correction term in the energy expansion.     

\section{Gross-Pitaevskii Theory}
\label{sec:gp}  
The GP equation for a condensate admits formulations in terms of the Hamilton least action principle with the action functional given as,
\begin{equation}
\ {S}=\int {L}dt \qquad\mbox{and}\qquad \ {L}=\int {\cal L}dr.
\label{eq:action}
\end{equation}
Here $\cal L$ is Lagrangian density and is given by, 
\begin{equation}
{\cal L}=\frac {i\hbar}{2} \left(\psi\frac{ \partial\psi^*}{\partial t}-\psi^*\frac{ \partial\psi}{\partial t}\right)+\frac{\hbar^2} {2 m}\left|\nabla\psi\right|^2+
\ {V}_{ext} \rho(r)+\varepsilon(\rho)\rho,
\label{eq:ld} 
\end{equation}              
where $V_{ext}$ is trapping potential, $m$ is mass of the atom, $\varepsilon(\rho)$ is self interaction energy, $\psi$ is condensate wave function and $\rho$ is density($\left|\psi\right|^2$). The normalization condition is, 
\begin{equation}
\int \left|\psi\right|^2 \vec{dr} = \int \rho\, \vec{dr} = {N}\quad 
\mbox{(total number of atoms).}
\label{eq:wavefun}
\end{equation}
By evaluating Lagrangian for a trial wave function, the solution of Lagrange's equations gives the parameters used in the trial wavefunction.

In the large gas parameter regime, we use the perturbative expansion ~\cite{arup2} for the interaction energy $\varepsilon(\rho)$ beyond mean field approximation by including the correction term calculated by Lee, Huang and Yang ( LHY) ~\cite{lee} as following,

\begin{equation}
\varepsilon(\rho)=\frac{2\pi\hbar^{2}a_{s}}{m}\rho[{1}+{C}_{1}(\rho {a}_{s}^{3})^{{1}/{2}}],
\label{eq:int_energy}	
\end{equation} 
where ${C}_{1}={128}/({15}{ \sqrt{\pi}})$. Here the first term in the expression of interaction energy $\varepsilon(\rho)$ represents the mean-field energy whereas the second term represents the LHY correction term. It is also known that this expansion is valid upto $ \rho a_{s}^{3} \approx 10^{-2} $ ~\cite{giorgini,arup2}.

Further, we assume that the Bose condensate is trapped in a harmonic potential of the form, 

\begin{equation}
\ {V}_{ext} = {\frac{1}{2}} m (\omega^{2}_{x}{x}^{2}+
\omega^{2}_{y}{y}^{2}+\omega^{2}_{z}{z}^{2}),
\label{eq:harm_pot}
\end{equation}
where $\omega_{x},\omega_{y},\omega_{z}$ are the trap frequencies along x, y and z-axes respectively. 

In the Q2D regime of confinement, the condensate is supposed to be confined in a trap which has frequency along one axis (z-axis) much larger than the frequencies along the other two axes (x-axis and y-axis). This implies that $\omega_{z}>>\omega_{x},\omega_{y}$ and motion in z-direction is frozen.  However, the oscillator lengths ${a}_{i}=(\hbar/m \omega_{i})^{1/2}, (i=x,y,z)$ along the three axes of the trap are still much larger than the scattering length so that the collisions are three dimensional. For the scattering length $a_s$ larger or comparable to the oscillator length along the confining axis, the scattering properties become significantly different from those of a three-dimensional case ~\cite{petrov1,petrov2,kolom,lieb,burnett}. We also assume that the chosen parameters for the condensate are such the nonlinear energy is smaller than the zero point energy implying that $({N}{a}_{s}/{a}_{z})<( {R}_{0}/{a}_{z})^{2}$, where ${R}_{0}$ is the size of the cloud in the x-y plane. Throughout the studies we have restricted $N$ and $a_s$ so that the above conditions are satisfied. Under these conditions we can write the condensate wave function $\psi$ in a factorized form as following ~\cite{kam},

\begin{equation}
\psi(\vec{r},t)= \phi(z)\Psi(x,y,t).
\label{eq:wf_xyt}
\end{equation}
We take the wave function along z-axis as 
\begin{equation}
\phi(z)={\frac{1}{(\sqrt{\pi}{a}_{z})^{{1}/{2}}}} \exp\left[{-}{z}^{2}/2{a}_{z}^{2}\right]
\label{eq:wf_z}
\end{equation}
such that 
\[\int \left|\phi\right|^2 dz={1}.
\]
 In order to study the time evolution of a Bose condensate confined in a Q2D trap as described above, we choose the  general trial wavefunction  in x-y coordinates as,
\begin{equation}
\Psi(x,y,t)={C(t)}\exp\left[{\left({-}{\frac{a^{2}_{0}}{2}}[\alpha(t){X}^{2}+
\beta(t){Y}^2+2\gamma(t){X}{Y}]\right)}\right],
\label{eq:wf_var_xy}
\end{equation}
where  $\alpha(t)$, $\beta(t)$ and $\gamma(t)$ are time dependent complex variational parameters. We have taken the trap to be anisotropic in the x-y plane with anisotropy parameter $\lambda $ given by $\omega_{x} = \lambda\omega_{y} =\omega_{0}\sqrt{\lambda} $. In equation (\ref{eq:wf_var_xy}), ${X}(={x}/{a}_{0})$ and ${Y}(={y}/{a}_{0})$ are dimensionless coordinate variables with $ {a}_{0}=(\hbar/{m \omega_{0}})^{{1}/{2}}$. We can write the condensate parameters $\alpha(t)$, $\beta(t)$ and $\gamma(t)$ such that 
\[ 
\alpha(t)=\alpha_{1}(t)+i\alpha_{2}(t),\ \beta(t)=\beta_{1}(t)+i\beta_{2}(t),\ \mbox {and}\ \gamma(t)=\gamma_{1}(t)+i\gamma_{2}(t).
\]
Here  $\alpha_{1}(t)$ and $\beta_{1}(t)$ represent the inverse square of the widths along x and y-axes respectively. The imaginary parts $\alpha_{2}(t)$ and $\beta_{2}(t)$ account for the time dependent phases. The parameters $\gamma_{1}(t)$ and $\gamma_{2}(t) $ representing the cross term in the coordinates account for the scissors modes of collective oscillation ~\cite{odelin,margo}. From the normalisation condition we can find that $\left|{C(t)}\right|^2={N}\sqrt{D}/\pi$, where ${D}=\alpha_{1}\beta_{1}-\gamma^{2}_{1}$.  

By substituting equations (\ref{eq:ld})-(\ref{eq:wf_var_xy}) in equation (\ref{eq:action}), we can evaluate Lagrangian as

\begin{equation}
{L}={L}_{T}+{E}_{K}+{E}_{ho}+{E}^{(1)}_{int}+{E}^{(2)}_{int},
\label{eq:lag}	
\end{equation}

where 

\lefteqn{
{L}_{T}={N}\hbar\omega_{0}\frac{1}{4D}\left[-\frac{1}{\omega_{0}}
(\alpha_{1}{\dot{\beta}}_{2}+\beta_{1}{\dot{\alpha}}_{2}-{2}{\gamma_{1}}{\dot{\gamma}}_{2})\right],}\bigskip
\lefteqn{
{E}_{K}={N}\hbar\omega_{0}\frac{a^{2}_{0}}{4D}\left[{D}(\alpha_{1}+\beta_{1})+
\alpha_{1}(\beta^{2}_{2}+\gamma^{2}_{2})+\beta_{1}(\alpha^{2}_{2}+\gamma^{2}_{2})-
{2}\gamma_{1}\gamma_{2}(\alpha_{2}+\beta_{2})\right]+\frac{{N}\hbar\omega_{z}}{4},}\bigskip
\lefteqn{
{E}_{ho}={N}\hbar\omega_{0}\frac{1}{4D}\left[\frac{\alpha_{1}}{\lambda{a}^{2}_{0}}+
\frac{\lambda\beta_{1}}{{a}^{2}_{0}}\right]+\frac{{N}\hbar\omega_{z}}{4}, }\bigskip
\lefteqn{
{E}^{(1)}_{int}={N}\hbar\omega_{0}{\frac{1}{4}\,D}\frac{2\sqrt{2}}{\sqrt{\pi}}
\left(\frac{a_{s}}{{a}_{z}}\right)a^{2}_{0}{N}{D}^{3/2},}\bigskip
\lefteqn{
{E}^{(2)}_{int}={N}\hbar\omega_{0}{C}_{1}{a}^{2}_{0}.\left(2\pi{a}_{s}\right)
\left(\frac{2}{5\pi\sqrt{\pi}}.\frac{a_{s}}{a_{z}}\right)^{3/2}{N}^{3/2}{D}^{3/4}.}
\bigskip
To know the static properties of the condensate, we substitute the static values of different parameters as $\alpha_{1}=\alpha_{10}$, $\beta_{1}=\beta_{10}$, $\alpha_{2}=\beta_{2}=0$, $\gamma_{1}=\gamma_{2}=0$ in the Lagrangian in equation (\ref{eq:lag}). Then, from the Lagrange's equations of motion, we can obtain the following coupled equations for $\alpha_{10}$ and $\beta_{10}$ 
\begin{eqnarray}
{a}^{2}_{0}{\alpha}^{2}_{10}-\frac{\lambda}{a^{2}_{0}}+\frac{k_{1}}{2}
\beta^{1/2}_{10}\alpha^{3/2}_{10}+{\frac{3}{4}}{k}_{2}\beta^{3/4}_{10}\alpha^{7/4}_{10}={0},
\label{eq:stat_alpha}\\	
{a}^{2}_{0}{\beta}^{2}_{10}-\frac{1}{\lambda{a}^{2}_{0}}+{\frac{k_{1}}{2}}
\alpha^{1/2}_{10}\beta^{3/2}_{10}+{\frac{3}{4}}{k}_{2}\alpha^{3/4}_{10}\beta^{7/4}_{10}={0},
\label{eq:stat_beta}
\end{eqnarray}
where\newline
\lefteqn{
{k}_{1}=\frac{4}{\sqrt{2\pi}}{a}^{2}_{0}\left({\frac{{a}_{s}}{{a}_{z}}}\right){N}},

\lefteqn{{k}_{2}=4.{\frac{128}{15\sqrt{\pi}}}{a}^{2}_{0}\left(2\pi{a}_{s}\right)
\left({\frac{2}{5\pi\sqrt{\pi}}}.{\frac{a_{s}}{{a}_{z}}}\right)^{3/2}{N}^{3/2}.}
\bigskip

These equations are coupled nonlinear equations. The terms with $k_2$ as coefficient in the above expressions appear due to the incorporation of LHY correction in the energy $\varepsilon(\rho)$. For the known values of parameters ( ${N},{a}_{s},{a}_{z}$, $\omega_{0}$ and $\lambda$ ) for the condensate in a Q2D trap, the static size-parameters $\alpha_{10}$ and $\beta_{10}$ along x and y-axes can be evaluated by solving the above equations. We have solved these equations numerically. The values of $\alpha_{10}$ and $\beta_{10}$ obtained thus have been used for the calculation of collective mode frequencies as described in the following section.

\section{Collective Modes}
\label{sec:cm}
The low energy excitations of condensate correspond to the small  oscillations in the cloud around the equilibrium configuration. Therefore we can expand the time dependent variational parameters around their static values in the following manner,

\begin{eqnarray}
\alpha_{1}=\alpha_{10}+\delta\alpha_{1}(t),\quad \beta_{1}=\beta_{10}+\delta\beta_{1}(t),\quad
\gamma_{1}=\delta\gamma_{1}(t)  \qquad (\gamma_{10}={0}),\nonumber\\
\alpha_{2}=\delta\alpha_{2}(t),\qquad \beta_{2}=\delta\beta_{2}(t),\qquad
\gamma_{2}=\delta\gamma_{2}(t)
\label{eq:var_par} 
\end{eqnarray}

After substituting the above values in the Lagrangian in equation (\ref{eq:lag}) and using the Euler-lagrange equations, the time evolution of the parameters $\delta\alpha_{1}$, $\delta\beta_{1}$ and $\delta\gamma_{1}$ can be obtained as
\begin{eqnarray}
\ddot{\delta\alpha_{1}}+\omega^{2}_{0}{a}^{2}_{0}\left[\frac{4\lambda}{{a}^{2}_{0}}-{\frac{k_1}{2}}\alpha_{10}\sqrt{\alpha_{10}\beta_{10}}-{\frac{3}{8}}{k}_{2}\alpha_{10}(\alpha_{10}\beta_{10})^{3/4}\right]\delta\alpha_{1} \nonumber\\
+\frac{\omega^{2}_{0}a^{2}_{0}}{2}\frac{\alpha^{2}_{10}}{\beta_{10}}\left[k_{1}(\alpha_{10}\beta_{10})^{1/2}+{\frac{9}{4}}{k}_{2}(\alpha_{10}\beta_{10})^{3/4}\right] \delta\beta_{1}=0, \label{eq:alp}\\
\ddot{\delta\beta_{1}}+\omega^{2}_{0}{a}^{2}_{0}\left[{\frac{4}{\lambda{a}^{2}_{0}}}-{\frac{k_1}{2}}{\beta_{10}}\sqrt{\alpha_{10}\beta_{10}}-
{\frac{3}{8}}{k}_{2}\beta_{10}(\alpha_{10}\beta_{10})^{3/4}\right]\delta\beta_{1} \nonumber \\
+\frac{\omega^{2}_{0}a^{2}_{0}}{2}\frac{\beta^{2}_{10}}{\alpha_{10}}\left[k_{1}(\alpha_{10}\beta_{10})^{1/2}+{\frac{9}{4}}{k}_{2}(\alpha_{10}\beta_{10})^{3/4}\right]\delta\alpha_{1}=0, \label{eq:bet}\\
\ddot{\delta\gamma_{1}}+\omega^{2}_{0}\left[\left(\lambda+{\frac{1}{\lambda}}\right)+2{a}^{4}_{0}\alpha_{10}\beta_{10}\right]\delta\gamma_{1}={0}. \label{eq:gam}
\end{eqnarray}
In the above equations (\ref{eq:alp}) - (\ref{eq:gam}), the first two equations (i.e. (\ref{eq:alp}) and (\ref{eq:bet})) are the coupled equations for $\delta\alpha_{1}$ and $\delta\beta_{1}$ . These equations represent the monopole and quadrupole modes of collective oscillations. The equation (\ref{eq:gam}) for $\delta\gamma_{1}$ is not coupled to the other two and represents the scissors mode of collective oscillation ~\cite{ghosh}. In order to calculate the the mode frequencies, we take the harmonic time dependence of amplitudes as 
\begin{eqnarray}
\delta\alpha_1=\delta\alpha_1(\omega)\exp(-{i}\omega{t}),
\ \delta\beta_1=\delta\beta_1(\omega)\exp(-{i}\omega{t}),\nonumber\\
\delta\gamma_1=\delta\gamma_1(\omega)\exp(-{i}\omega{t}).
\label{eq:harm_varpar}
\end{eqnarray}  

This results in the solution for mode frequencies as 
\begin{equation}
\omega_\pm ^{2} =\left({\frac{a_1}{2}}+{\frac{b_2}{2}}\right)\pm\left[
\left({\frac{a_1}{2}}-{\frac{b_2}{2}}\right)^2 + a_2\,b_1\right]^{1/2}
\label{eq:monoquad}
\end{equation}
and
\begin{equation}
{\frac{\omega^2_{s}}{\omega^2_0}}=\left(\lambda+{\frac{1}{\lambda}}\right)+2{a}^4_0(\alpha_{10}\beta_{10}),
\label{eq:scissor}
\end{equation}
where,

\lefteqn{a_1=\omega^{2}_{0}{a}^{2}_{0}\left[\frac{4\lambda}{{a}^{2}_{0}}-
{\frac{k_1}{2}}\alpha_{10}\sqrt{\alpha_{10}\beta_{10}}-
{\frac{3}{8}}{k}_{2}\alpha_{10}(\alpha_{10}\beta_{10})^{3/4}\right],}
\bigskip
\lefteqn{b_1=\frac{\omega^{2}_{0}a^{2}_{0}}{2}\frac{\alpha^{2}_{10}}{\beta_{10}}\left[k_{1}
(\alpha_{10}\beta_{10})^{1/2}+{\frac{9}{4}}{k}_{2}(\alpha_{10}\beta_{10})^{3/4}\right],}
\bigskip
\lefteqn{a_2=\frac{\omega^{2}_{0}a^{2}_{0}}{2}\frac{\beta^{2}_{10}}{\alpha_{10}}\left[k_{1}
(\alpha_{10}\beta_{10})^{1/2}+{\frac{9}{4}}{k}_{2}(\alpha_{10}\beta_{10})^{3/4}\right],}
\bigskip
\lefteqn{b_2=\omega^{2}_{0}{a}^{2}_{0}\left[{\frac{4}{\lambda{a}^{2}_{0}}}-
{\frac{k_1}{2}}\beta_{10}\sqrt{\alpha_{10}\beta_{10}}-
{\frac{3}{8}}{k}_{2}\beta_{10}(\alpha_{10}\beta_{10})^{3/4}\right].}
\bigskip

\begin{figure}[htbp]
\resizebox{0.75\columnwidth}{!}{%
 \includegraphics{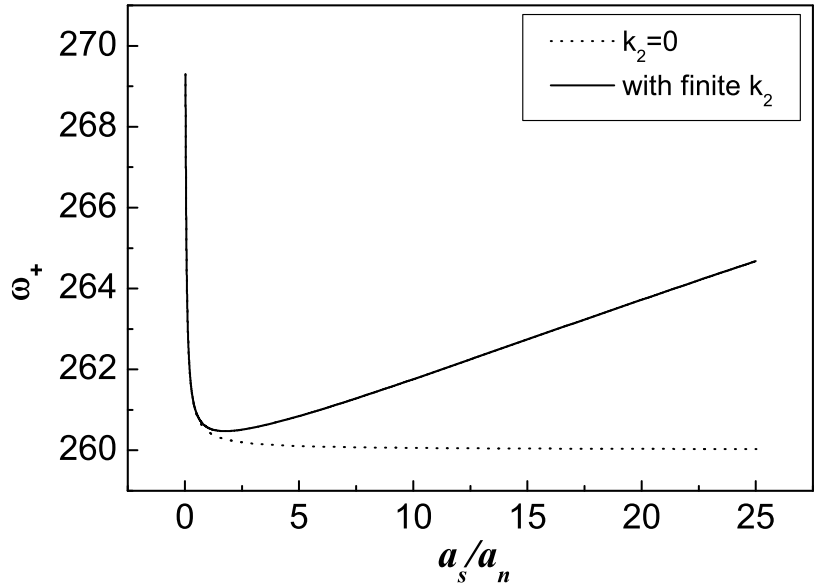} 
}
\caption{Variation of monopole mode frequency with scattering length. The other parameters are $ N = 10^4$, $ \omega_z = 2\pi \times 1000Hz $, $\omega_0 = 2\pi \times 20 Hz$, $ a_n = 27.5 A^0$ and $\lambda = 0.8$. The dotted curve shows frequency values calculated within the mean-field theory (i.e. $k_{2}=0$) whereas the continuous curve shows that after incorporating LHY correction term (i.e. $k_{2}=finite$).}
\label{fig:mono}
\end{figure}

Here $\omega_{+} $ is monopole mode frequency, $\omega_{-} $ is quadrupole mode frequency and $\omega_{s} $ is scissors mode frequency. The terms with $k_{2} $ as coefficient in the above expressions of frequencies represent the corrections due to LHY term in the interaction energy $\varepsilon(\rho)$. It can be noted that monopole and quadrupole mode frequencies depend explicitly upon $k_{2} $  whereas scissors mode frequency does not depend upon it.  Variation of monopole mode frequency $\omega_{+} $ with scattering length (i.e. $a_{s}/a_{n} $, $a_{n} $ is scattering length in absence of magnetic field) is shown in FIG. \ref{fig:mono}. It is evident that with finite value of $k_{2} $, monopole mode frequency changes with $a_{s}/a_{n} $ in a differnt manner than that with $k_{2} =0 $. The frequency values from the two curves differ in quantitative as well as in qualitative manner as shown in the figure. After the initial fall in the frequency $\omega_{+}$ with increasing $a_{s}/a_{n} $, it becomes nearly constant if $k_{2} =0$. However for a finite value of $k_{2}$, after the initial fall, there is monotonous increase in $\omega_{+}$ values with increasing $a_{s}/a_{n} $. Although the modifications in the values of $\omega_{+}$ with finite $k_{2}$ are just by a few percent, nonetheless, they can be easily detected as has been pointed out earlier ~\cite{string}. 

\begin{figure}[htbp]
\resizebox{0.75\columnwidth}{!}{%
 \includegraphics{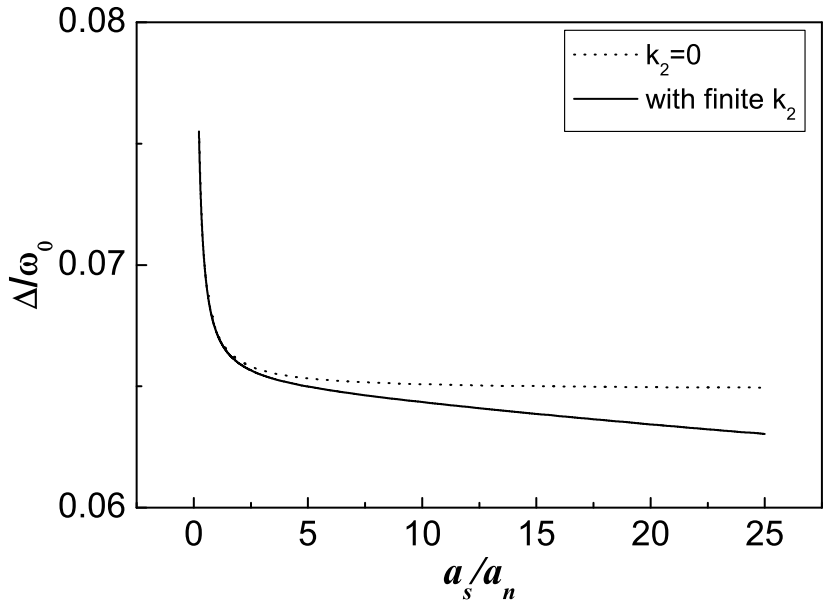} 
}
\caption{Variation of mode splitting ( $\Delta=\omega_{s} -\omega_{-}$ ) with $a_s/a_n$. The other parameters are $ \omega_z = 2\pi \times 1000Hz $, $ \omega_0 = 2\pi \times 20 Hz$,  $N=10^4$, $ a_n = 27.5 A^0$ and $\lambda=0.8$. The dotted curve shows $\Delta$ values calculated within the mean-field theory (i.e. $k_{2}=0$) whereas the continuous curve shows that with incorporating LHY term in the inteaction energy (i.e. $k_{2}=finite$).} 
\label{fig:scissor}
\end{figure}

 As has been discussed earlier ~\cite{ghosh} the splitting between quadrupole and scissors modes occurs in a 2D anisotropic trapped Bose condensate. The value of mode splitting ($\Delta=\omega_{s} -\omega_{-}$) depends upon anisotropy and interaction strength. To assess the effect of LHY term on the value of $\Delta$, we have calculated it as function of  $a_s/a_n$ for finite $k_{2} $ as well as for $k_{2} $ = 0. The results are as shown in the FIG. \ref{fig:scissor}. The values shown here are for anisotropy parameter $\lambda=0.8$. As is evident from the figure, the splitting $\Delta$ calculated with finite $k_{2}$  keeps on changing with $a_s/a_n$ whereas it becomes nearly constant with $a_s/a_n$ for $k_{2}=0 $. It is worth mentioning that the scissors mode frequency $\omega_{s} $ does not explicitly depend upon the LHY term ( i.e. $k_2$). However, its dependence on this term occurs via parameters $\alpha_{10}$ and $\beta_{10}$, which is quite small. Therefore the difference in the $\Delta$ values in two curves in FIG.\ref{fig:scissor} arises mainly due to dependence of quadrupole frequency $\omega_{-} $ on the terms containing $k_2$.  

\section{Conclusion}
\label{sec:conc}
 We have theoretically studied the collective mode frequencies of a quasi two-dimensional Bose condensate in the large gas parameter regime. We find that in this regime, the inclusion of LHY correction term beyond the mean-field term in the interaction energy  leads to quantitative as well as qualitative changes in the dependence of collective mode frequencies on interaction strength. Such regime of parameters is experimentally accessible with the present advancements in this field and modifications in the mode frequencies are easily detectable.
 
\begin{acknowledgments}
We thank Dr S. C. Mehendale for a critical reading of the manuscript.

\end{acknowledgments}

\end{document}